\documentclass[10pt,conference,a4paper]{IEEEtran}
\usepackage{fancyhdr} 
\usepackage[USenglish,american]{babel}
\usepackage{epsfig,graphics,subfigure,graphicx,latexsym,longtable,amsmath,amscd,latexsym,amssymb,mathrsfs,syntonly,eucal}
\usepackage{multirow}
\usepackage[usenames]{color}
\usepackage[T1]{fontenc}
\usepackage{bm,cite}
\usepackage{amsbsy}
\usepackage{latexsym}
\usepackage{wasysym}
\usepackage{placeins}
\usepackage[lined,boxed,commentsnumbered]{algorithm2e}
\usepackage{lipsum}
\usepackage{url}
\usepackage{array}
\usepackage{tabu}
\SetKwInput{KwInput}{Input}
\SetKwInput{KwOutput}{Output}
\usepackage{booktabs,siunitx}
\usepackage{geometry}
 \newcommand\blfootnote[1]{%
  \begingroup
  \renewcommand\thefootnote{}\footnote{#1}%
  \addtocounter{footnote}{-1}%
  \endgroup
}

\usepackage{bm,cite}
\usepackage{amsmath}
\usepackage{amsbsy}
\usepackage{latexsym}
\usepackage{amssymb}
\usepackage{wasysym}
\usepackage{mathtools}

\DeclareMathAlphabet{\mathbit}{OML}{cmr}{bx}{it}
\DeclareMathAlphabet{\mathsf}{OT1}{cmss}{m}{n}
\DeclareMathAlphabet{\mathTXf}{OT1}{cmss}{bx}{it}

\DeclareMathOperator{\Transpose}{T}

\DeclareMathOperator*{\argmax}{arg\ max}

\DeclareMathAlphabet{\mathpzc}{OT1}{pzc}{m}{it}

\newcommand{\Tr}{{\Transpose}}

\newcommand{\He}{{{\text{H}}}}

\newcommand{\eqdef}{=\vcentcolon}

\graphicspath{{./figures/}}

\title{Round Trip Time Estimation
Utilizing Cyclic Shift of Uplink Reference Signal}

\author{
\IEEEauthorblockN{Rajeev Gangula\IEEEauthorrefmark{1}, Tommaso Melodia\IEEEauthorrefmark{1}, Rakesh Mundlamuri\IEEEauthorrefmark{1}\IEEEauthorrefmark{2}, and Florian Kaltenberger\IEEEauthorrefmark{1}\IEEEauthorrefmark{2}
}
\IEEEauthorblockN{\IEEEauthorrefmark{2}Communication Systems Department,
EURECOM, Biot, France
}

\IEEEauthorblockN{\IEEEauthorrefmark{1}Institute for the Wireless Internet of Things, Northeastern University, Boston, USA 
}
}

\usepackage{tikzpagenodes,etoolbox}
\usetikzlibrary{calc}
\usepackage[contents={}]{background}
\AddEverypageHook{%
\ifnumequal{\thepage}{1}{%
    \tikz[remember picture,overlay]{%
        \node[draw,
        minimum width=1.03\textwidth,
        text width=1.02\textwidth,
        font=\scriptsize
        ]
        at ($(current page header area) - (0,5pt)$)
        {%
        This work has been submitted to the IEEE for possible publication.\\
        Copyright may be transferred without notice, after which this version may no longer be accessible.
        };
    }
}{}
}

\begin{document}

\maketitle

\begin{abstract}

In the context of fifth-generation new radio (5G NR) technology, it is not possible to directly obtain an absolute uplink (UL) channel impulse response (CIR) at the base station (gNB) from a user equipment (UE). The UL CIR obtained through the sounding reference signal (SRS) is always time-shifted by the timing advance (TA) applied at the UE. The TA is crucial for maintaining UL synchronization, and transmitting SRS without applying the TA will result in interference. In this work, we propose a new method to obtain absolute UL CIR from a UE and then use it to estimate the round trip time (RTT) at the gNB. This method requires enhancing the current 5G protocol stack with a new Zadoff-Chu (ZC) based wideband uplink reference signal (URS). Capitalizing on the cyclic shift property of the URS sequence, we can obtain the RTT with a significant reduction in overhead and latency compared to existing schemes.
The proposed method is experimentally validated using a real-world testbed 
based on OpenAirInterface (OAI).
\end{abstract} 

\blfootnote{This work is supported by OUSD (R\&E) through Army Research Laboratory Cooperative Agreement Number W911NF-24-2-0065. The views and conclusions contained in this document are those of the authors and should not be interpreted as representing the official policies, either expressed or implied, of the Army Research Laboratory or the U.S. Government. The U.S. Government is authorized to reproduce and distribute reprints for Government purposes notwithstanding any copyright notation herein.}
\section{Introduction} \label{sec:intro}
Apart from offering high data rates and low latency communication, next-generation networks
are expected to provide precise positioning and environment sensing capabilities \cite{bourdoux2020,rel_17_design}.
The Third Generation Partnership Project (3GPP)
has defined several positioning methods and procedures in the fifth-generation new radio (5G NR) technology and expected to enhance them in the future releases, with sixth-generation (6G) standardization starting from the year 2025 \cite{ericson}. Timing-based positioning methods in 5G include downlink time difference of arrival (DL-TDoA), uplink time difference of arrival (UL-TDoA), and multi-cell round trip time (multi-RTT) \cite{DwivediEtal_21,italiano2024tutorial}.

The multi-RTT positioning method involves estimating a user equipment’s (UE’s) location using the round trip time (RTT) between the UE and multiple base stations (gNBs). 
The UE's 2D position can be estimated with the RTT from 
at least three well-placed gNBs using the trilateration procedure. Unlike the TDoA method, very tight (order of nanoseconds) synchronization among gNBs is not required in RTT methods. 
The 3GPP has defined the following RTT-based positioning methods: Enhanced 
cell ID (ECID) type I \& II and multi-RTT \cite{rodeWhte,DwivediEtal_21,italiano2024tutorial,3gpp2018nr_36_355,3gpp2018nr_36_305,3gpp2018nr_38_305,3gpp2018nr_37_355}.

Despite being standardized, the performance evaluation of the 3GPP-compliant
positioning methods has traditionally been limited to either system-level simulations
or real-world experimental evaluations in proprietary settings \cite{DwivediEtal_21,italiano2024tutorial,qualcomm1,qualcomm2}. 
However, with the success and wide adoption of open-source 5G reference implementations such as 
OpenAirInterface (OAI), it is now possible for researchers to validate their positioning
algorithms with real-world experiments.
Few works have demonstrated the timing-based 5G positioning techniques in real-world experiments using the OAI platform \cite{ahadi20235g,malik2024conceptreality5gpositioning,del2022proof,del2023first,mundlamuri2024novel,mundlamuri20245gnrpositioningopenairinterface}. While the work in \cite{mundlamuri2024novel} is based on RTT, the works in \cite{ahadi20235g,malik2024conceptreality5gpositioning,del2022proof,del2023first} focus on DL-TDOA  and UL-TDoA methods.

This work introduces a new RTT estimation scheme
for 6G systems. 
The proposed method enhances the current 5G protocol stack with a
new uplink reference signal (URS), and significantly reduces
the overhead and latency in obtaining the RTT compared to
existing schemes in the literature and 3GPP standards \cite{mundlamuri2024novel,rodeWhte,DwivediEtal_21,italiano2024tutorial,3gpp2018nr_36_355,3gpp2018nr_36_305,3gpp2018nr_38_305,3gpp2018nr_37_355}.
Specifically, our contributions are

\begin{itemize}
    \item A new reference signal, URS, similar to the
sounding reference signal (SRS) in 5G is introduced that
allows us to estimate the RTT at the gNB
    \item The OAI gNB and UE protocol stack is updated with the URS feature
    \item The proposed RTT method is then validated with over-the-air experiments
    \item The framework allows us to jointly estimate the RTT from multiple URS measurements, resulting in better performance in low signal-to-noise ratio (SNR) scenarios.
\end{itemize}
We now proceed with the background that lays down the basis to understand the state-of-the-art RTT methods and the novelty of the proposed scheme.
\section{Background}\label{sec:backg}
This section provides a brief overview of the tools and procedures necessary for RTT measurement in 5G NR.

\subsection{5G Synchronization}\label{sec:5gsync}

In 5G NR, the synchronization process consists of downlink (DL) and uplink (UL) synchronization.
The DL synchronization is achieved by UE acquiring the frame and symbol boundary by decoding the synchronization signal block (SSB) transmitted by the gNB.
The UE can then decode the master information block (MIB) from the physical broadcast channel (PBCH) and system information block (SIB) from the physical downlink shared channel (PDSCH). The parameters extracted from the MIB and SIB are necessary to perform UL synchronization.

\begin{figure}[t]
\centerline{\includegraphics[width=3in]{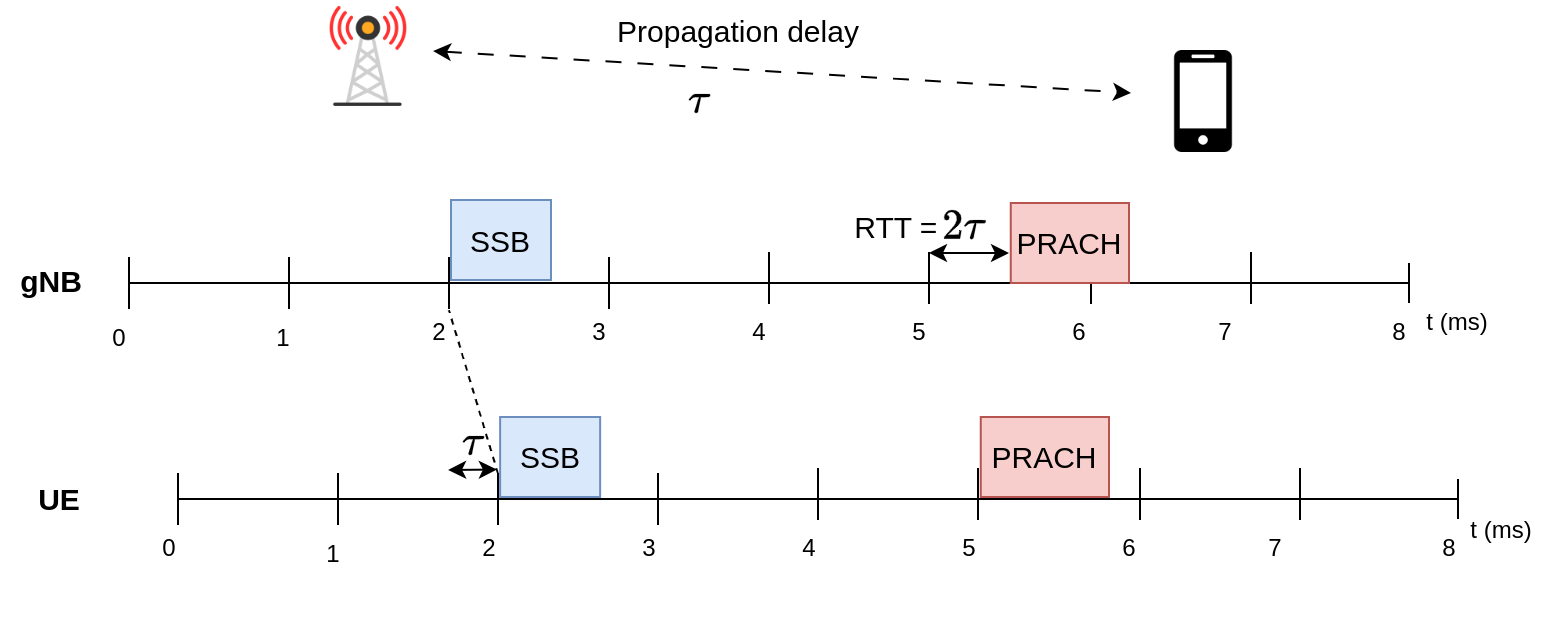}}
\caption{RTT estimation with RACH}
\label{fig:dl_ul_sync}
\end{figure}

\begin{figure}[t]
\centerline{\includegraphics[width=2.8in]{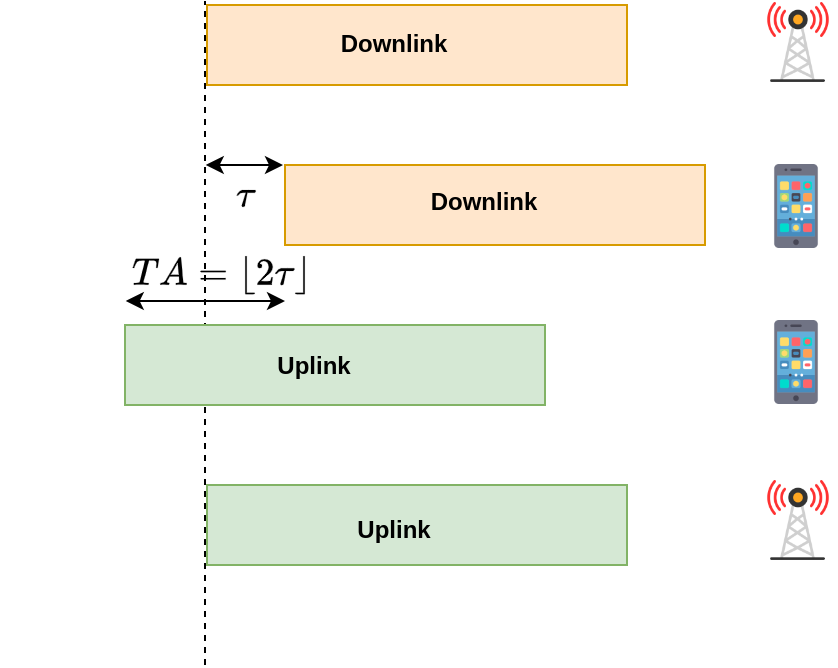}}
\caption{UL timing adjustment at UE}
\label{fig:ta}
\end{figure}

The gNB handles multiple UE's situated at different distances across the cell. Therefore, each UE’s UL transmission time needs to be adjusted to align with the gNB’s UL reception.
UL synchronization is achieved through a combination of the random access (RA) procedure and timing advance (TA) loops. The RA procedure is utilized during the initial access or when the UE loses UL synchronization. The gNB can issue TA adjustment commands to the UE when its in a connected state.
The steps in UL synchronization are as follows:

\begin{itemize}
    \item The UE selects a preamble from a set of predefined preambles (for contention-based) or uses a preamble configured by the gNB (contention-free) and transmits it over the physical random access channel (PRACH).
    \item By detecting the preamble, the gNB can measure the coarse RTT.
    A quantized version of the RTT, known as TA, is then sent to the UE via the random access response (RAR).
    \item When the UE successfully receives the RAR, it adjusts its UL timing by the TA value.
    \item Although the initial TA is sent via the RAR, gNB can periodically send updated UL timing corrections to the UE via TA commands. These TA commands are crucial in maintaining UL synchronization in the event of UE mobility or clock drift.
\end{itemize}
The synchronization process is shown in Figures \ref{fig:dl_ul_sync} and \ref{fig:ta}.


\subsection{Reference Signals}\label{sec:5grefsig}

Several reference signals are used for channel estimation and positioning in 5G NR.
Specifically, positioning reference signals (PRS) in the DL and sounding reference signals (SRS) in the UL
are essential for many 5G NR positioning methods \cite{DwivediEtal_21}.
These reference signals were introduced in 4G and extended to 5G, offering better resolution and accuracy \cite{3gpp2018nr_38_211}.
While the SRS is generated using the Zadoff-Chu sequence, PRS is generated using quadrature phase shift keying (QPSK) modulated 31-length gold sequence \cite{3gpp2018nr_38_211}.
The OAI gNB and UE (nrUE) support both SRS and PRS.
Detailed instructions for configuring the PRS in OAI can be found in \cite{prs}.


\subsection{Zadoff-Chu sequences}\label{sec:zcseq}

Zadoff-Chu (ZC) sequences are widely used in 5G NR due to a number of desirable properties. 
A ZC sequence of length $N_{\text{ZC}}$, which must be an odd number, and root $q \in [1,2,\ldots,N_{\text{ZC}}-1]$ is defined as
\begin{equation}
    x_q[n] = \exp \left(-j\frac{\pi q n (n+1)}{N_{\text{ZC}}} \right), 0 \leq n \leq N_{\text{ZC}}-1.
\end{equation}
Some key properties of the ZC sequences are given below.
\begin{itemize}
    \item All the elements in a ZC sequence have unit amplitude.
    \item Normalized cyclic auto-correlation: When the root $q$ is relatively prime to $N_{\text{ZC}}$, 
    \begin{equation}\label{eq:aucorr}
 \frac{1}{N_{\text{ZC}}}\sum_{n=0}^{N_{\text{ZC}}-1}x_q[n] x^{*}_q[(n+\nu) \text{mod} N_{\text{ZC}}] = \delta[\nu],
    \end{equation}
where mod represent the modulo operation and $\delta[\nu]$ represents the Kronecker delta function.
There are $N_{\text{ZC}}$ unique cyclic shifts of the sequence $x_q[n]$.
    \item Normalized cyclic cross-correlation: when $|q_1-q_2|$ and $N_{\text{ZC}}$ are relatively prime,
   \begin{equation}\label{eq:crcorr}
 \frac{1}{N_{\text{ZC}}} \sum_{n=0}^{N_{\text{ZC}}-1}x_{q_1}[n] x^{*}_{q_2}[(n+\nu) \text{mod} N_{\text{ZC}}] = \frac{1}{\sqrt{N_{\text{ZC}}}}.
    \end{equation}
    \item The discrete Fourier transform (DFT) (or its inverse) of a ZC sequence is also a ZC sequence. 
\end{itemize}

\section{Prior art} \label{sec:part}
The existing approaches for obtaining RTT in the cellular networks, termed as
\textit{Scheme A}, \textit{Scheme B}, and \textit{Scheme C} are presented in this section.
While \textit{Scheme A} and \textit{Scheme B} are in 3GPP standards; \textit{Scheme C} is a recently proposed enhancement. 

\noindent \textbf{\textit{Scheme A}:}
RTT can be obtained at the gNB from PRACH during the RA procedure. This process is outlined in Section \ref{sec:5gsync} and illustrated in Figure \ref{fig:dl_ul_sync}. 

The RTT accuracy of this approach is limited by the low bandwidth of the RACH signal. Moreover, the RA procedure is performed only during the initial access or in the event of UL
synchronization failure.

\begin{figure}[t]
\centerline{\includegraphics[width=2.8in]{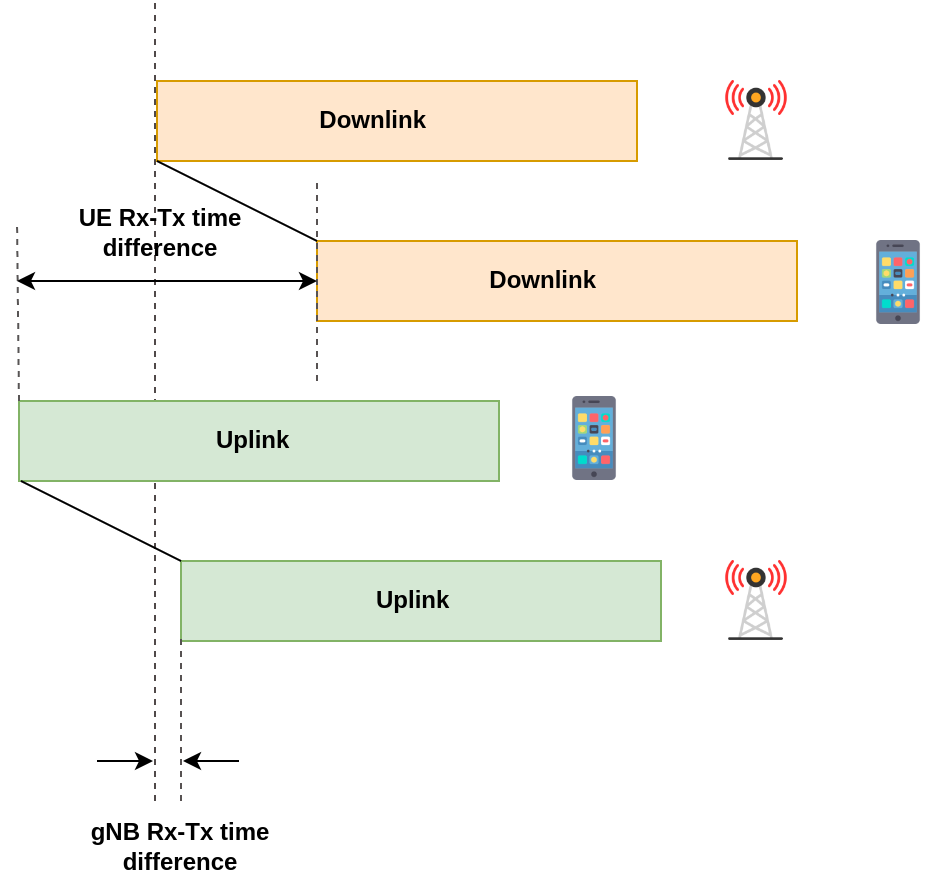}}
\caption{Estimation of RTT in \textit{Scheme B}.}
\label{fig:rtt_sys}
\end{figure}

\noindent \textbf{\textit{Scheme B}:}   
Using wideband reference signals like SRS and PRS, UE and gNB can estimate the receive-transmit timing difference (Rx-Tx time difference) measurements. Figure \ref{fig:rtt_sys} depicts this method. The RTT is calculated as
   \begin{align*} 
       \text{RTT} = \text{UE Rx-Tx }&\text{time difference} \\
       &+\\
       \text{gNB Rx-Tx } & \text{time difference}.  
    \end{align*} 
    The ECID type I scheme \cite{3gpp2018nr_36_355,3gpp2018nr_36_305} and multi-RTT scheme in 5G NR \cite{3gpp2018nr_38_305,3gpp2018nr_37_355} follow this method.
    The details of Rx-Tx time difference measurement reporting is provided in \cite{3gpp2018nr_38_133}.
    
\textit{Scheme B}'s RTT estimation is more accurate than \textit{Scheme A}'s, as we use wideband reference signals. However, this performance comes at the expense of increased latency and resource consumption, as the UE must send its Rx-Tx time difference measurements to the network.

\begin{figure}[t]
\centerline{\includegraphics[width=2.8in]{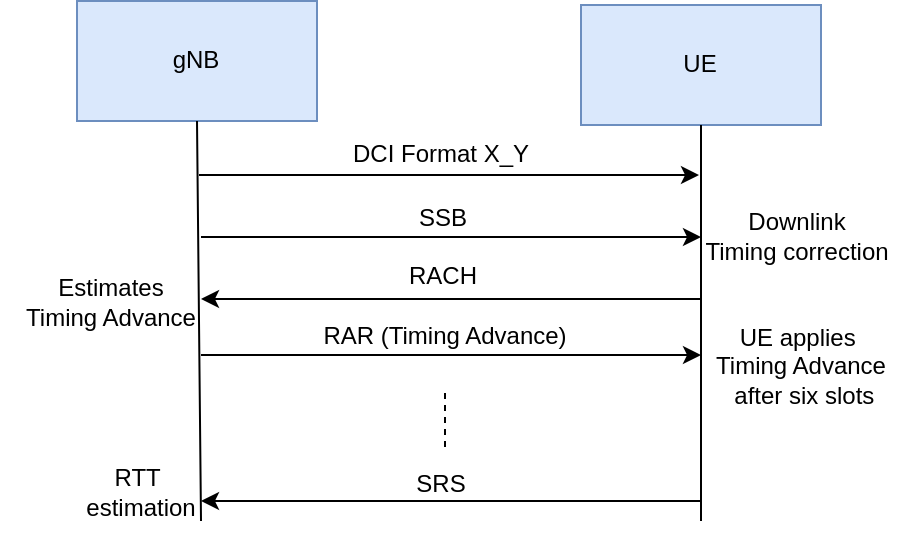}}
\caption{Signaling mechanism in \textit{Scheme C} \cite{mundlamuri2024novel}.}
\label{fig:pdcch_order}
\end{figure}

\noindent \textbf{\textit{Scheme C}:}
Recently, the authors in \cite{mundlamuri2024novel} proposed a RTT estimation scheme by combining the TA measurement obtained from PRACH and SRS channel measurements. This method considerably improves the RTT estimation performance compared to \textit{Scheme A} while not incurring extra UE Rx-Tx time difference measurement reporting costs as in \textit{Scheme B}. Moreover, RTT can be obtained even when the
5G UE is in a radio resource control (RRC) inactive state.
The signaling scheme is presented in Figure \ref{fig:pdcch_order}.



The drawback of \textit{Scheme C} lies in the fact that a contention-free RA procedure is required to estimate the RTT, resulting in additional radio resource consumption and increased latency.

In this work, a novel RTT method that overcomes the shortcomings of the above mentioned state-of-the-art schemes is presented. 


\section{Detailed Description} \label{sec:detdesc}

The proposed method relies on PRS and URS to obtain the RTT.
The central idea is to use the current TA and the delay (in samples) estimated from PRS at the UE as a cyclic shift while generating the URS. We present the generation and reception of the URS followed by the signaling procedure.

\subsection{Uplink reference signal} \label{sec:schmc}
The URS is generated from a base ZC sequence whose properties are outlined in Section \ref{sec:zcseq}.
The URS base sequence of length $N_{\text{ZC}}$, with root $q \in [1,2,\ldots,N_{\text{ZC}}-1]$ being relative prime to $N_{\text{ZC}}$
is defined as
\begin{equation}
    x_q[n] = \exp \left(-j\frac{\pi q n (n+1)}{N_{\text{ZC}}} \right), 0 \leq n \leq N_{\text{ZC}}-1.
\end{equation}
Let the cyclic shifted sequence is denoted by $x_{q,\nu}[n] \eqdef x_q[(n-\nu) \text{mod} N_{\text{ZC}}]$, where $\nu$ is the cyclic shift.
The following property plays an essential role in the proposed method,
\begin{align*} 
  x_q[n]  &\overset{\mathcal{F} }{\longrightarrow} X_q[k]\\
  x_{q,\nu}[n] &\overset{\mathcal{F} }{\longrightarrow} X_q[k] \exp{\left(-j\frac{2 \pi k \nu}{N_{\text{ZC}}}\right)},
\end{align*} 
where $\mathcal{F}$ denotes the Fast Fourier Transform (FFT) operation and $0 \leq k \leq N_{\text{ZC}}-1$. Note that there are $N_{\text{ZC}}$ unique cyclic shifts.

\subsection{Transmit and Receive Chain}\label{sec:txrx}
\begin{figure}[t]
\centerline{\includegraphics[width=3.4in]{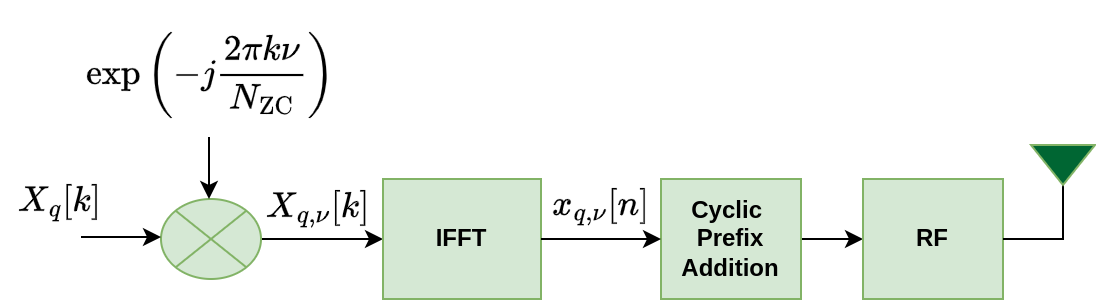}}
\vspace{3mm}
\caption{URS transmission at UE.}
\label{fig:tx_block}
\end{figure}

The transmitter block in an orthogonal frequency-division multiplexing (OFDM) based system, as in 5G NR, is shown in Figure \ref{fig:tx_block}. The time-domain URS signal generated at the UE is denoted by $x_{q,\nu}[n]$.

\begin{figure}[t]
\centerline{\includegraphics[width=3.4in]{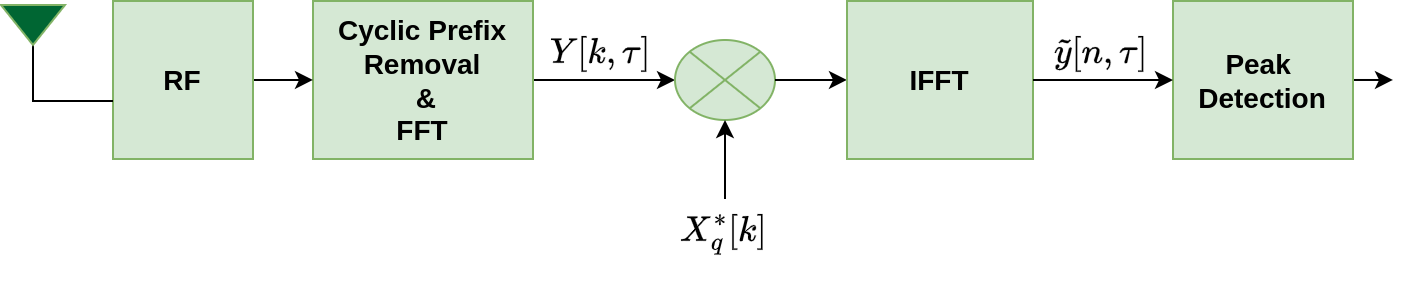}}
\caption{URS receiver at gNB.}
\label{fig:rx_block}
\end{figure}

The received time domain baseband signal at the gNB can be written as
$$
y[n,\tau] = x_{q,\nu}[n] \circledast h[n-\tau]+w[n], 0 \leq n \leq N_{\text{ZC}}-1
$$
where $h[.]$ denotes the wireless channel inducing a propagation delay $\tau$, $\circledast$ denotes the circular convolution and $w[.]$ is the additive white Gaussian noise (AWGN).
The receiver block at the gNB is shown in Figure \ref{fig:rx_block}.
After performing FFT
$$
Y[k,\tau] = X_{q,\nu}[k] H[k] e^{-j\frac{2\pi k \tau}{N_{\text{ZC}}}} +W[k], 0 \leq k \leq N_{\text{ZC}}-1,
$$
where $H[k]$ and $W[k]$ are the frequency domain channel and AWGN at $k$-th sub-carrier, respectively.
By doing the correlation with the base reference signal sequence,
$$
Y[k,\tau] X^{*}_{q}[k] = H[k] e^{-j\frac{2\pi k \tau}{N_{\text{ZC}}}} X_{q,\nu}[k] X^{*}_{q}[k] +\Tilde{W}[k].
$$
$$
= H[k] e^{-j\frac{2\pi k (\tau+\nu)}{N_{\text{ZC}}}}+\Tilde{W}[k].
$$
After doing IFFT
$$
\tilde{y}[n,\tau]= \mathcal{F}^{-1}\left\{Y[k,\tau] X^{*}_{q}[k]\right\} = h[n-(\tau+\nu)]+\tilde{w}[n]
$$


The peak of the CIR $|\tilde{y}[n,\tau]|$ at $\tau+\nu$ corresponding to the cyclic shift $\nu$ and propagation delay $\tau$.







\subsection{{\textit{Cyclic shift based RTT}}} \label{sec:schme}

\begin{figure}[t]
\centerline{\includegraphics[width=3in]{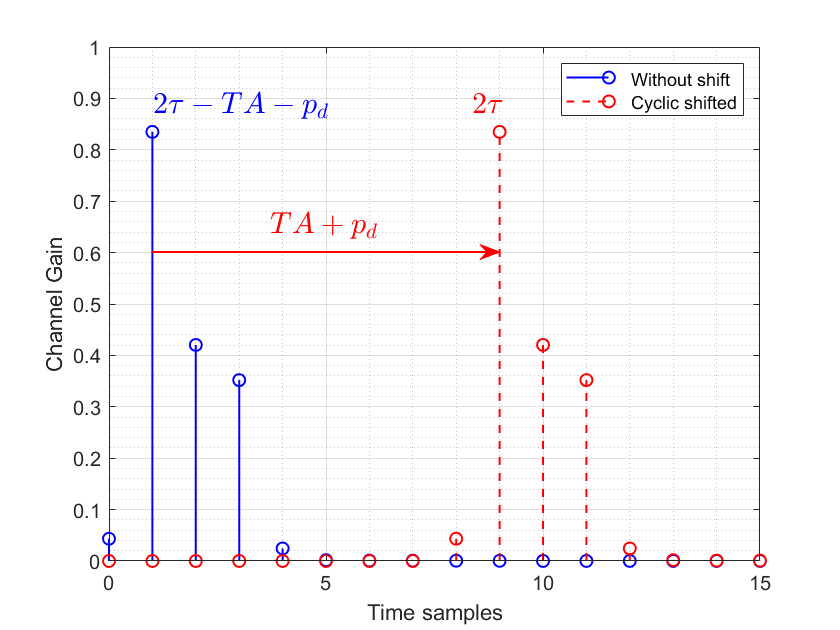}}
\vspace{-2mm}
\caption{Cyclic shifted URS CIR.}
\label{fig:cs_shift}
\end{figure}

\begin{figure}[t]
\centerline{\includegraphics[width=3in]{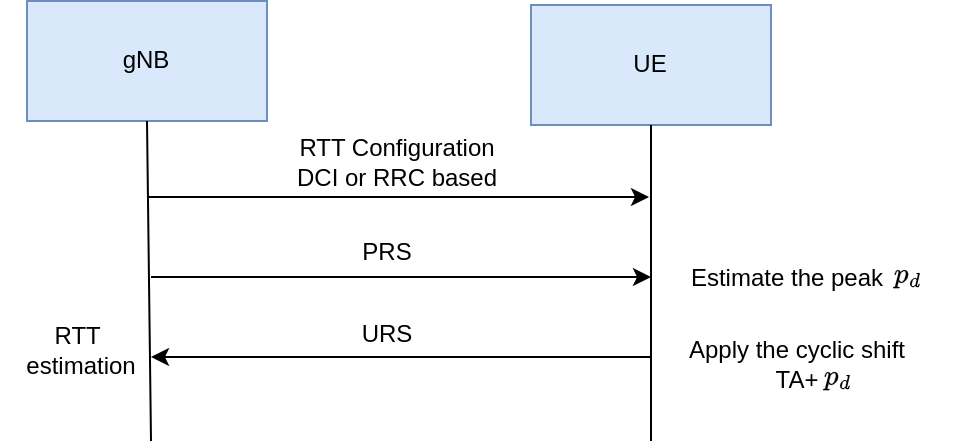}}
\caption{Signaling for cyclic shift-based RTT Scheme.}
\label{fig:e_signaling}
\end{figure}

Leveraging the properties of the URS, we now present our cyclic shift based RTT estimation method.
The signaling scheme is presented in Figure \ref{fig:e_signaling} and summarized as follows:

\begin{itemize}
    \item The gNB configures the URS and PRS resources and prepares the UE by sending the downlink Control information (DCI) through the physical downlink control channel (PDCCH) \cite{mundlamuri2024novel} or by RRC signaling \cite{3gpp2018nr_38_305}.
    \item UE will measure the channel impulse response (CIR) first peak, $p_d$ samples, derived from the PRS
    \item The cyclic shift of $\text{TA} + p_d$ is applied on the base URS sequence $x_q[n]$ resulting in $x_q[(n-\text{TA}-p_d) \text{mod} N_{\text{ZC}}]$.
    \item UE transmits the URS
    \item The delay estimated from URS derived CIR first peak at the gNB will give
       \begin{align*} 
       \text{RTT} &= 2\tau-\text{TA}-p_d + \underbrace{\text{TA}+p_d}_\textrm{Cyclic shift}  \\
       &=  2\tau.
    \end{align*}
This is illustrated in Figure \ref{fig:cs_shift}.
\end{itemize}

Note that the cyclic shift must satisfy $0\leq\text{TA}+p_d< N_{\text{ZC}}$.

In the next section, we discuss the extensions to the proposed scheme to multiple users, and when the TA value is greater than the URS sequence length.

\section{Extensions} \label{sec:extn}

\subsection{Larger Timing Advance} \label{sec:longta}
For large TA values, the proposed scheme
 can be extended by using multiple base URS sequences having different roots. For example, let us consider a scenario where $0\leq \text{TA}+p_d < 2N_{\text{ZC}}$.
 Here the gNB can configure the UE with two URS base sequences $x_{q_1}[n]$ and $x_{q_2}[n]$.
The URS generation at UE is given by
\begin{itemize}
    \item if $0 \leq \text{TA}+p_d <N_{\text{ZC}}$, generate $x_{q_1}[n-(\text{TA}+p_d)~ \text{mod} N_{\text{ZC}}]$ 
    \item if $N_{\text{ZC}} \leq \text{TA}+p_d <2 N_{\text{ZC}}$, generate $x_{q_2}[n-(\text{TA}+p_d-N_{\text{ZC}})~ \text{mod} N_{\text{ZC}}]$,
\end{itemize}

The roots are selected such that
$q_1$, $q_2$ and $|q_1-q_2|$ are relative prime with $N_{\text{ZC}}$.
The gNB can detect the delay and the root index using the auto and cross correlation properties stated in \eqref{eq:aucorr} and \eqref{eq:crcorr}.
If the URS peak, denoted by ${p}_{u}$, is detected at the gNB by doing correlation 
of the received signal with 
\begin{itemize}
    \item base sequence $x_{q_1}[n]$, then $\text{RTT} = {p}_u $
    \item base sequence $x_{q_2}[n]$, then $\text{RTT} = N_{\text{ZC}} + {p}_u $
\end{itemize}

\subsection{Multiple Users} \label{sec:multiUE}
Multiple UEs can be multiplexed in an OFDM symbol using
the comb structure in the frequency domain similar to SRS \cite{3gpp2018nr_38_211}.
This way, we can estimate the RTT from multiple UEs using
the same OFDM symbol.

The detailed implementation and performance analysis of
these extensions are out of scope and are left for future work.

\begin{figure*}
    \centering    \includegraphics[width=0.8\textwidth]{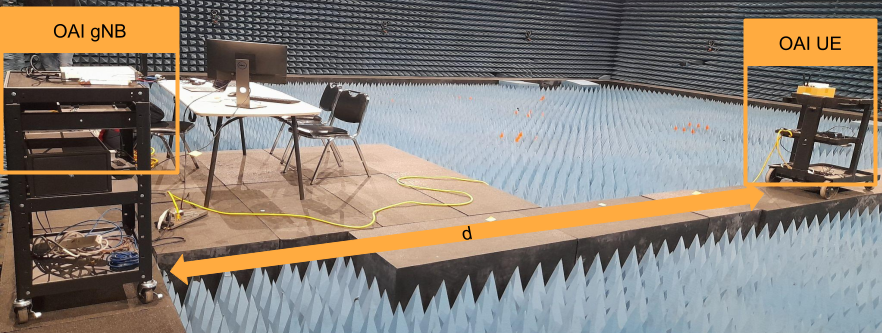}
    \caption{Experimental setup in the anechoic chamber.}
    \label{fig:chamber}
\end{figure*}
\section{Experimental Setup}\label{sec:Expsetup}

To experimentally validate the proposed method, we consider a scenario
where a single antenna gNB and a UE communicate over a line-of-sight channel.
The gNB and the UE rely on the OAI 5G NR protocol
stack \cite{KalaloAbhiLuh_20} and USRP B210 software-defined boards.
Moreover, the SC2430 NR signal conditioning module is used as an external RF front-end at the gNB \cite{sc_mod}. 
The experiment is performed in an anechoic chamber at the Northeastern University Innovation Campus at Burlington, as shown in Figure~\ref{fig:chamber}.
The gNB and UE operate in 5G NR band n78 with the system parameters listed in Table~\ref{tab:1}.
We use the phy-test mode of the OAI \cite{phy_test}.This mode allows us to run the OAI UE and gNB physical layer procedures while abstracting the higher layer protocol stack.
The hardware delays are calibrated and later compensated in the channel measurements.
We have enhanced the OAI software stack to implement the proposed scheme presented in Figure \ref{fig:e_signaling}.

\noindent \textbf{URS support in OAI:} 
While the PRS is already implemented in OAI, we have added the URS feature in OAI nrUE and gNB.
The URS implementation follows Figures \ref{fig:tx_block} and \ref{fig:rx_block}.
The URS base sequence is a ZC sequence of length $N_\text{ZC}$, with $N_\text{ZC}$ being prime and $N_\text{ZC} < K$, where
$K$ denotes the FFT size. While mapping into the OFDM symbol, the remaining $N_\text{ZC}-K$ resource elements (REs) are filled with zeros. Due to this up-sampling effect, the cyclic shift applied to the URS is calculated as
\begin{equation}\label{eq:csupsamp}
    \nu = \lceil(\text{TA}+p_d)\times K/N_\text{ZC}\rceil,
\end{equation}
where $\text{TA}$ (in samples) is the current timing advance at the nrUE and $p_d$ is the peak (in samples) detected from the PRS channel estimate. 

The PRS is transmitted on 12 symbols of slot 1 of every frame by the gNB, and
URS is transmitted periodically in symbol 13 of slot 8 by the nrUE\footnote{Note that this can be made flexible and left for future implementation.}.
The UE applies the cyclic shit (based on the received PRS in slot 1 and TA) calulated in \eqref{eq:csupsamp} on the URS sequence trasnmitted in slot 8.
The TDD slot configuration is listed in Table~\ref{tab:1}.
With these modifications in the OAI stack, we can experiment with the proposed framework.

\begin{table}[htbp]
\caption{System Parameters}
\centering
\begin{tabular*}{\columnwidth}{@{\extracolsep{\fill}}cc}
\toprule
Parameters & Values \\
\midrule
TDD slot configuration & DL DL DL DL DL DL DL Mixed UL UL\\
System bandwidth & 38.16 MHz \\
Subcarrier Spacing ($\Delta f$) & 30 KHz \\
Centre frequency ($f_c$) & 3.69 GHz \\
Sampling rate ($f_s$) & 46.08 MHz \\
FFT size ($K$) & 1536 \\
URS bandwidth & 37.77 MHz\\
URS length ($N_{\text{ZC}}$) & 1259\\
PRS bandwidth & 37.44 MHz\\
PRS symbols & 12\\
PRS Comb & 2\\
\bottomrule
\end{tabular*}
\label{tab:1}
\end{table}

\section{RESULTS}\label{sec:result}

The performance of the proposed RTT estimation scheme is evaluated 
in terms of 
empirical cumulative distribution function (CDF) of the range estimation error in low and high UL SNR scenarios.
The CDF is obtained from 48,000 channel measurements collected using the signaling procedure illustrated in Figure \ref{fig:e_signaling} and the setup described in Section~\ref{sec:Expsetup}.
These measurements are obtained by fixing the position of the gNB and moving the UE in a straight line between 3 to 10 meters with a 1-meter increment. At every point and SNR, a total of 
6,000 measurements are collected. The experiment is performed in an anechoic chamber, as shown in Figure~\ref{fig:chamber}.

We apply two algorithms, namely, the peak detector (PD) and the matched filter (MF) on 
the collected URS channel estimates to obtain the empirical CDF.
The range estimated using PD and MF is given by
\begin{equation}
\hat{d}\ (\textrm{PD}) = \frac{\text{c}}{2f_s}\times \frac{1}{M}\sum_{m=1}^{M} \argmax \left|\tilde{\bm{y}}_m\right|,
\end{equation}

\begin{equation}
    \hat{d}\ (\textrm{MF}) = \frac{c}{2} \times \argmax_{\hat{\tau}} \bm{v}(\hat{\tau})^{\He} {\bf{R_y}} \bm{v}(\hat{\tau}),
\end{equation}
where 
$$
{\bf{R_y}} = \frac{1}{{M}}\sum^{{M}}_{m=1}{\bm{Y}}_m {\bm{Y}}_m^{\He},
$$
$\tilde{\bm{y}}_m=\big[\tilde{y}_m[0,\tau],\tilde{y}_m[1,\tau],\dots,\tilde{y}_m[K-1,\tau]\big],m\in[1,M]$ is the $m$-th estimated URS CIR, ${\bm{Y}}_m=\big[{Y}[0,\tau],{Y}[1,\tau],\dots,{Y}[K-1,\tau]\big]^{\Tr}$ is the $m$-th estimated URS channel frequency response, $M$ is the number of measurements, $\bm{v}(\hat{\tau})=
[1,e^{-j2\pi \Delta f \hat{\tau}},\ldots,e^{-j2\pi (K-1)\Delta f \hat{\tau}}]^\Tr$, $f_s$ is the sampling rate and $c$ is the speed of light. 

\begin{figure}[t]
\centerline{\includegraphics[width=3in]{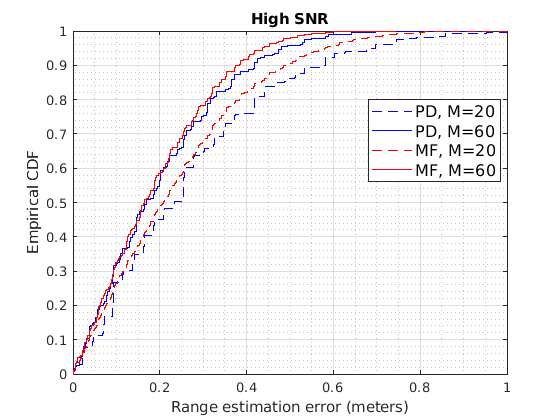}}
\caption{CDF of the range
estimation error.}
\label{fig:high_snr}
\end{figure}

\begin{figure}[t]
\centerline{\includegraphics[width=3in]{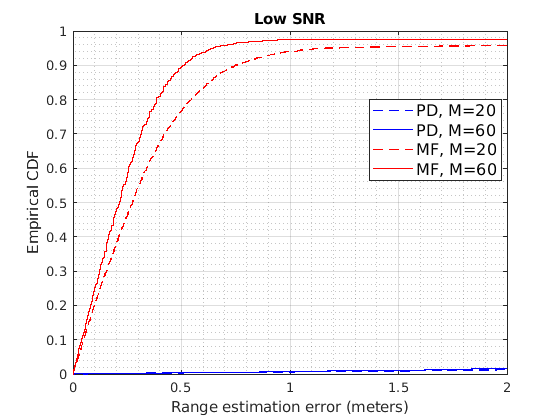}}
\caption{CDF of the range
estimation error.}
\label{fig:low_snr}
\end{figure}

The CDF of the range estimation error for MF and PD algorithms at high and low UL SNR is shown in Figures~\ref{fig:high_snr} and~\ref{fig:low_snr}. 
In the case of high UL SNR, the USRP TX gain of the UE is set to $89.5$ dB, resulting in an estimated UL SNR of $30$ dB, while in the case of a low SNR, the UE TX gain is reduced by $50$ dB. The gNB USRP TX gain is fixed throughout the experiments.
While the MF and PD schemes have similar performance at high SNR, the MF algorithm has significantly better performance in low SNR scenario. In the low SNR scenario, for $M{=}20$ measurements, the range estimation error of MF is below 0.8 meters for 90\% of the time. Furthermore, by increasing the number of measurements from $M{=}20$ to $60$, we see an improvement in the performance for both methods.

\section{Conclusion}\label{sec:conclusion}

This work has introduced a novel mechanism that facilitates the estimation of RTT between a UE and gNB.
Our method comes with a reduction in overhead and latency compared to the existing schemes. The proposed method relies on PRS and a ZC sequence-based wideband URS.
We have
enhanced the OAI 5G protocol stack with the URS feature
and validated the proposed scheme with real-time over-the-air experiments.
Future work includes extending to multiple users and large timing advance values. Because of its simplicity and reduction in radio resource overhead, we believe it is a good candidate for positioning and sensing in 6G systems.


\bibliographystyle{IEEEtran}
\bibliography{References}

\end{document}